\documentclass[aps,prl,twocolumn,showpacs,floats]{revtex4}
\usepackage{graphicx}%
\usepackage{epsfig}
\usepackage{amsmath}%
\usepackage{amssymb}%

\def\beq{\begin{equation}}

\def\eeq{\end{equation}}

\def\e{\epsilon}

\def\mx{\mathrm{max}}

\def\pr{{\sl Phys. Rev.}\ }

\def\prl{{\sl Phys. Rev. Lett.}\ }

\begin{document}

\bibliographystyle{prsty}


\title{\Large\bf Theory of Magneto-resistance of Disordered Superconducting Films }

\author{Yonatan Dubi$^1$, Yigal Meir$^{1,2}$ and Yshai Avishai$^{1,2}$}

\affiliation{
$^{1}$ Physics Department, Ben-Gurion University, Beer Sheva 84105, Israel\\
$^{2}$ The Ilse Katz Center for Meso- and Nano-scale Science and
Technology, Ben-Gurion University, Beer Sheva 84105, Israel }
\date{\today}

\begin{abstract} \noindent Recent experimental studies of magneto-resistance in
disordered superconducting thin films reveal a huge peak (about 5
orders of magnitude). While it may be expected that magnetic field
destroys superconductivity, leading to an enhanced resistance,
attenuation of the resistance at higher magnetic fields is
surprising.
 We propose a model which accounts for the
experimental results in the entire range of magnetic fields, based
on the formation of superconducting islands due to fluctuations in
the superconducting order parameter amplitude. At strong magnetic
fields Coulomb blockade in these islands gives rise to negative
magneto-resistance. As the magnetic field is reduced the effect of
Coulomb blockade diminishes and eventually the magneto-resistance
changes sign. Numerical calculations show good qualitative
agreement with experimental data. \end{abstract}

\pacs{71.30.+h,73.43.-f,73.43.Nq,74.20.Mn} \maketitle The
interplay between superconductivity and disorder is a
long-standing problem, dating back to the late fifties \cite{Anderson,Gorkov}, resulting
in the common wisdom that weak disorder has no dramatic effect on
superconductivity. Strong disorder, however, may have a profound
effect, driving the system from a superconducting (SC) to an
insulating state. Such a SC-insulator transition (SIT) was
observed in two-dimensional amorphous superconducting films
\cite{Goldman_review}. Reducing the film thickness or an
increasing perpendicular magnetic field drives these films (which
are held below their bulk critical temperature)
from a SC state, characterized by a vanishing resistance as $T
\rightarrow T_c$, to an insulating state, characterized by a
diverging resistance as $T \rightarrow 0$. The possibility of
tuning the system continuously between these two phases is
a manifestation of a quantum phase transition
\cite{QPTreview}. \par The origin of this transition is still in
debate. While some theories \cite{dirtybosons} claim that it may
be understood in terms of Cooper-pair scattering out of the SC
condensate into a Bose-glass state (so-called "dirty boson"
models), there is evidence, both experimental
\cite{percolation1,Hebard,kapitulnik,russians} and theoretical
\cite{percolation2}, that a percolation description of the SIT is
more adequate for, at least, some of these samples. 
\par More insight into the nature
of the SIT may be gained by looking at the
magneto-resistance (MR) on its insulating side. Decade-old experiments observed
non-monotonic MR, exhibiting a shallow peak at some magnetic field $B_{\mx}$
\cite{Ovadyahu,Gent}. Recent experiments \cite{Murty}
show, however,  that in some samples the effect is dramatic, with the resistance value at the peak, $R_{\mx}$,
 reaching as high as a few orders of magnitude its value at the SIT. As the
magnetic field is further increased the MR drops back a few orders
of magnitude (inset of Fig.~3). Further investigations of this effect
\cite{Kapitulnik} reveal that disorder also has a major influence.
With increasing disorder strength, the
critical field for the SIT, $B_{c}$, diminishes, while $R_{\mx}$ increases.
 The temperature dependence of the MR at high temperatures fits an
activation-like behavior, $R \propto \exp (T_0 / T) $, with a
magnetic field dependent $T_0$ as seen in the inset of Fig.~4(a).
 At lower temperatures there is a
deviation from this behavior towards some weaker temperature dependence. While
an enhancement of the resistance with increasing magnetic field may be understood in terms
of decreasing SC order, the suppression of the resistance with further increase of magnetic field
remains a puzzle.
 \par Here we propose a model for the MR in
the entire range of magnetic fields. The model is based on three
assumptions. The first is that disorder induces formation of SC
islands (SCIs) due to fluctuations in the amplitude of the SC
order parameter. This concept has already been used to interpret
some experiments \cite{Ovadyahu, Murty}. It was directly observed
in other systems by STM measurements \cite{STM} and further
corroborated by numerical simulations \cite{ghosal}. The second
assumption is that as the magnetic field is increased, the
concentration and size of these SCIs  decrease. Preliminary numerical results 
support this picture and will be reported in the future. The third assumption is that the SCIs have 
a charging energy, and thus, a Cooper pair entering the SCI via an
Andreev tunneling process, has to overcome the charging energy.
The charging energy is expected to be inversely proportional to
the island size, and thus to increase with increasing magnetic
field. 

\par In order to see the mechanism by which the MR can be
negative, consider such a system in the strong magnetic field
regime, $B>>B_{\mx}$ (Fig.~1(a)). Due to the strong magnetic field
the SCIs are small and have a large charging energy. There are two
types of trajectories available for electron transport: those
which follow normal areas of the sample ("normal paths", solid
lines in Fig.~1(a)) and those in which an electron tunnels into a
SCI via the Andreev channel ("island paths", dashed lines in
Fig.~1(a)). The resistance of the normal paths has some value
(which may depend on e.g. length, temperature, etc.) and is
assumed to be only weakly affected by magnetic field. Due to the
Coulomb blockade, transport through the SCI paths is thermally
activated, and hence the resistance of the island paths is of the
form $R \sim \exp (E_{c}/T)$, where $E_c$ is the charging energy
of the island. If $E_c$ is large then the main contribution to the
conductance is due to transport along the normal paths. Consistent
with experiment, the MR in this regime is small.
\par As the magnetic field is decreased (but still in
the regime $B>B_{\mx}$, Fig.~1.(b)), more SCIs are created and
their size increases, but they are still small enough such that
transport along normal paths is favorable. However, some paths which were
normal at higher fields (e.g. bottom solid line in Fig.~1(a)) now
become island paths and hence unavailable for electron transport
(bottom dashed line in Fig.~1(b)). Thus, the effective phase space available
for electron transport diminishes, resulting
in a negative MR. Eventually, at a certain magnetic field $B=B_{\mx}$
(Fig.~1(c)) some SCIs are large  enough so that their charging
energy is small and the resistance through them is comparable to
the resistance through normal paths. At this point the resistance
reaches its maximum value, since as the magnetic field is further decreased
(Fig.~1(d)) the SCIs are so large that transport through them is
always preferred over transport through normal paths. Increase in number and size
of the SCIs will thus result in a decrease in the resistance. At the critical 
field $B_{c}$ the SCIs percolate through
the system, resulting in an insulator-to-SC
transition. 
We note that if the temperature is smaller than the Josephson coupling between two islands then
SC correlations will extend between them, effectively joining them to a single island. Thus, we expect that
 at low temperatures the geometry  will be temperature-dependent, resulting in a temperature-dependent critical point
 as is indeed seen in some experiments.

\begin{figure}[h] \centering
\includegraphics[width=8.2truecm]{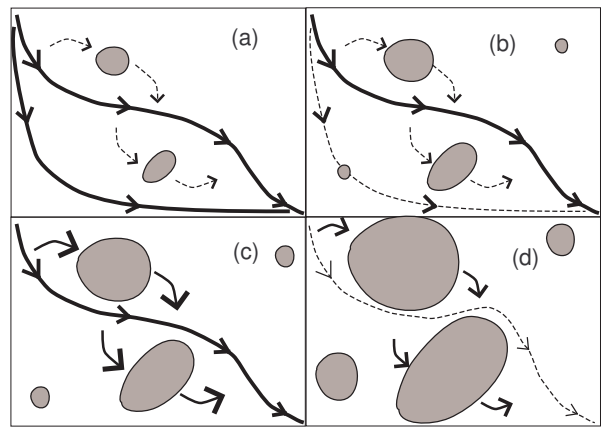} \caption{\footnotesize
Schematic representation of the model :(a) At strong magnetic fields,
$B>>B_{\mx}$, The system is composed of small superconducting
islands with large charging energy. In this regime transport
through normal paths (solid lines) is always preferable than
transport through the superconducting islands (SCIs) (see text). (b) As the magnetic field is
decreased, but still $B>B_{\mx}$, more SCIs appear, resulting in a
decrease in available trajectories for transport (bottom solid
line in (a) and bottom dashed line in (b))  and hence negative
magneto-resistance. (c) At a certain field $B_{\mx}$ the resistance
of normal paths and paths that include SCIs is comparable,
resulting in a peak in the resistance. (d) For even lower fields,
$B_{c}<B<B_{\mx}$, transport through SCIs is always favored,
resulting in positive magneto-resistance.} \end{figure}

\par In order to substantiate these heuristic arguments we model the thin film by a square lattice, 
where each site can be either normal, with
probability $p$, or SC, with probability $1-p$, corresponding to the
concentration of normal areas and SCIs in the sample. The probability
$p$ is a function of the magnetic field $B$, and we may assume that $p(B)$ is an increasing monotonic
function. To describe the disordered system in strong magnetic fields, in the
absence of SCIs, we follow Ref. \cite{VRH} and assign a resistance
between any two normal sites of the form
\beq
R_{NN} =R_0 \exp \left( \frac{2 r_{ij}}{ \xi_{loc}}+\frac{|\e_{i}|+|\e_{j}|+|\e_{i}-\e_{j}|}{2kT} \right),
\label{VRH resistance}
 \eeq
where $R_0$ is a constant, $r_{ij}$ is the distance between sites $i$ and $j$,
$\xi_{loc}$ is the localization length, $\e_{i}$ is the energy of
the $i$-th site measured from the chemical potential (taken from a
uniform distribution $[-W/2,W/2] $) and $T$ is the temperature.
The localization length $\xi_{loc}$ is taken to be small (in units of
lattice constant), effectively allowing only nearest- and next-nearest-neighbor hopping.
 All SC sites that are linked to each other are considered a single SCI.
The resistance between two (neighboring) SC sites, $R_{SS}$, is taken to be very small compared with Eq.(\ref{VRH resistance}),
but still not zero and temperature dependent, in such a way that it vanishes as $T\rightarrow 0$ (distant SC sites
are disconnected). The calculations of resistance were conducted with several functional forms for $R_{SS}(T)$ (power law, exponential dependence, etc.) and no qualitative difference between them was found. The resistance between a
normal site and a SC site (local N-S junction) is taken to be \beq R_{NS} \propto \exp
(E_{c}/kT), \label{G_NS} \eeq where $E_c$ is the charging energy of the
island. For simplicity, and to avoid additional parameters, the charging energies of all islands were taken to be the same, independent of island size. It is demonstrated below that the  main experimental observations
can be well understood even under such an assumption  \cite{E-scales}. 

The lattice is connected to electrodes (Fig.~2) and
the resistance is calculated numerically using Kirchhoff's laws.
 The left-most and right-most
links in Fig.~2 are taken to be SC, thus avoiding a strong dependence of the resistance on the properties
of the edge sites.
\begin{figure}[h] \centering
\includegraphics[width=7truecm]{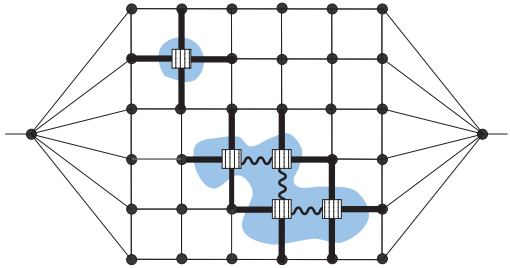} \centering
\caption{\footnotesize The lattice model is composed of regular sites and
superconducting sites. Clusters of the latter form SCIs (shaded islands). The resistance between two normal sites
(small circles) is given by Eq.~\ref{VRH resistance}. The resistance between two SC sites (filled squares) is
very small (and becomes smaller with decreasing temperature, see
text) for neighboring sites (wavy lines) and is infinite (no
link) between non-neighboring sites. The resistance between
neighboring normal and SC sites (thick lines) is much higher than
the resistance between two normal sites, and is exponential in the
charging energy of the SC island (Eq.~\ref{G_NS}).} \end{figure} In Fig.~3 we plot the calculated
resistance (on a log scale) as a function of the probability $p$
for different temperatures. The calculations were conducted on
lattices of size $25 \times 25$ (and repeated for
different lattice sizes, with no qualitative difference in the
results) and log-averaged over 100 realizations of disorder. A peak in the MR is observed at $p_{\mx}=0.5$, with peak resistance four orders of magnitude larger than the resistance at the transition.
The results are compared with the experimental data of
\cite{Murty} (inset of Fig.~3), and good qualitative agreement is
evident. Notice that the critical probability, $p_c$, defined as the probability at which the resistance is temperature-independent,
is shifted from the percolation critical probability. The reason for this is that the resistance of the SC links, $R_{SS}$, is finite. As the temperature is decreased the critical probability moves
 towards the percolation critical probability, eventually reaching it at $T=0$.

\begin{figure}[h] \centering
\includegraphics[width=9truecm]{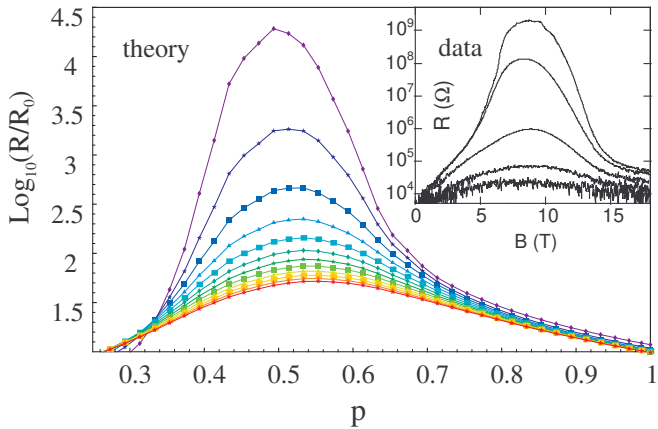} \centering
\caption{\footnotesize Numerical results : resistance (on a
log scale) as a function of probability $p$ for different
temperatures, with the parameters $W=0.4,~E_c=4,~\xi_{loc}=0.1,~T=0.1,0.2,...,4 $.
This is to be (qualitatively) compared with the
experimental data (inset) of \cite{Murty}.} \end{figure}

\par When fitting the resistance as a function of temperature to
an activation-like behavior, $R \propto \exp (T_0 / T) $, we find
a non-monotonic dependence of $T_0$ on the probability $p$
(Fig.~4), resembling the experimental data of \cite{Murty} (upper
inset of Fig.~4) and of \cite{Kapitulnik}. The activation
temperature rises from $T_0 \approx  W$ at $p =1$ to $T_0 \approx
E_c$ for $p=p_{\mx}$. It then drops back again due to increasing
weight of SC areas in the sample, eventually reaching $T_0=0$ at
the transition. Thus, in our picture $T_0$ is determined by the
Coulomb blockade energy and not by the SC gap, as was initially
suggested in \cite{Murty}.
\par  At high temperatures the activation fit is excellent
for all values of $p$. For low temperatures, on the other hand,
the fit becomes worse (lower inset of Fig.~4). Similar results were presented
 in the experimental data
of \cite{Murty}. The reason for this is that at low enough temperatures tunneling into the SCIs is suppressed, except  very close to the SIT. Since the resistance through normal areas is activated not by $E_c$ but rather by $W$, the slope of the Arrhenius plot changes, as is also evident in the experiment. We thus predict that the low temperature behavior near $B_c$ will be the same as in higher temperatures in strong magnetic fields $B>>B_{\mx}$.
\begin{figure}[h] \centering 
\includegraphics[width=9truecm]{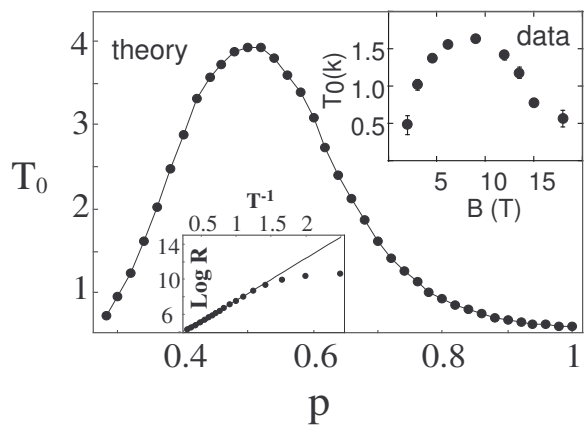} \caption{\footnotesize
The activation energy $T_0$ as extracted from fitting the
resistance to an activation behavior, obtained from the numerical
calculation and from the experimental data of \cite{Murty}
(inset). Lower inset : an Arrhenius plot of the resistance as a function of temperature for $p=p_{\mx}=0.5$. A deviation from an activated behavior is clearly seen.}
\end{figure}
\par The fact that the temperature dependence is not a pure activation in the whole magnetic field and
 temperature range is crucial to the observation of such a peak in the MR. If one assumes solely a magnetic field dependent activation energy, $T_0 (B)$, then for different temperatures $T_1$ and $T_2$ the ratio
\beq
\frac {\log (R(B,T1)/R_0)}{\log (R(B,T2)/R_0)} =\frac{T_1/T_0(B)}{T_2/T_0(B)}=\frac{T_1}{T_2} ~~,
\eeq
would be independent of field, in contrast with the experimental observation.
\par

The amount of disorder affects our model in several ways. First,
the width of the energy distribution $W$ increases, though this
has a minor effect on the behavior near the MR peak. Second the
initial concentration of normal islands, $p_0$, increases with
disorder. If one assumes $p(B)\simeq p_0 + \alpha B^x$ (where $\alpha$ is some constant and $x$
is probably equal to $2$ for small fields, as the system is
symmetric under reversal of the magnetic field direction), then
$B_\mx=(p_\mx-p_0)^{1/x}/\alpha$, namely $B_{\mx}$ decreases with
increasing disorder. Lastly, the typical size of the SCIs
decreases, leading to an enhancement of the Coulomb charging energy
and an exponential increase of $R_{\mx}$. These latter two
points are consistent with the experimental observations
\cite{Kapitulnik}, though quantitative predictions require a
detailed study of the distribution of the SCIs with
disorder and magnetic field (currently underway).

As mentioned in \cite{Kapitulnik}, at very strong fields (up to 30
T) the resistance of some samples saturates at values somewhat
larger than the resistance just above the (temperature-driven)
superconducting transition. Our model provides a natural
explanation for this. At strong magnetic fields all vestiges of SC
correlations are gone, and one is left with an insulator (which
is hardly affected by magnetic field) at a lower temperature, and
thus with higher resistance.
\par The theory presented here may also account for the MR in the presence of a tilted magnetic field \cite{tilted}.
Taking into account the finite width of the sample, a tilted
magnetic field suppresses the SCIs both in-plane and in the "thin"
direction. Due to the finite width, the electron trajectories are
now 3-dimensional, and thus changes in the SCI size due to
parallel magnetic field affects the resistance in a similar way as
described above, leading to the same non-monotonic MR for parallel
fields. This also explains the shift in the location of MR peak
towards higher fields for parallel fields \cite{tilted} .

\par To summarize, we have demonstrated that competition between normal
electron and Cooper-pair transport, generated by Coulomb
blockade of superconducting islands and driven by perpendicular
magnetic field, may yield non-monotonic magneto-resistance. This is accompanied by a change in the temperature dependence of the resistance, resulting from a crossover from nearest-neighbor hopping in normal areas to tunneling into the SC islands. This crossover may lead to a change of several orders of magnitude in the resistance, as seen experimentally.
\par We thank G. Sambandamurthy and D. Shahar for making their data
available to us, and A. Aharony for fruitful discussions. This
research has been funded by the ISF.


\begin{references}
\bibitem{E-scales}
Even in this approximation the reasoning described in the text is
valid, as a decrease in magnetic fields (at large fields,
$B>B_{\mx}$) leads to an increase in the number of local N-S
junctions and thus to an increase in the number of paths which are
characterized by a resistance given by (\ref{G_NS}). In this case
$B_{\mx}$ is the magnetic field where the normal resistance
associated with going around a typical island is equal to
$R_{NS}$.
\end{references}
\end{document}